\documentclass[aps,showpacs,twocolumn,superscriptaddress,nofootinbib,floatfix,section 10pt]{revtex4-2}
\usepackage{amsfonts,amssymb,stmaryrd,latexsym,amsmath,braket}
\usepackage{graphicx,subfigure}
\usepackage{comment}
\usepackage{times}
\usepackage{slashed}
\usepackage{bm}
\usepackage{braket}
\usepackage{appendix}
\usepackage{enumitem}
\usepackage{multirow}
\usepackage{array}
\usepackage{amsmath}
\usepackage[colorlinks=true,backref=true,linktocpage=true,citecolor=blue,urlcolor=blue,linkcolor=blue,pdfpagemode=UseOutlines]{hyperref}
\usepackage[dvipsnames]{xcolor}

\begin{document}
\title{Asymptotic errors in adiabatic evolution
}
\author{Thomas D. Cohen}
\email{cohen@umd.edu}
\affiliation{Department of Physics and Maryland Center for Fundamental Physics, University of Maryland, College Park, MD 20742 USA}

\author{Hyunwoo Oh}
\email{hyunwooh@umd.edu}
\affiliation{Department of Physics and Maryland Center for Fundamental Physics, University of Maryland, College Park, MD 20742 USA}

\begin{abstract}
The adiabatic theorem in quantum mechanics implies that if a system is in a discrete eigenstate of a Hamiltonian and the Hamiltonian evolves in time arbitrarily slowly, the system will remain in the corresponding eigenstate of the evolved Hamiltonian. 
Understanding corrections to the adiabatic result that arise when the evolution of the Hamiltonian is slow---but not arbitrarily slow---has become increasingly important, especially since adiabatic evolution has been proposed as a method of state preparation in quantum computing. This paper identifies two regimes, an adiabatic regime in which corrections are generically small and can depend on details of the evolution throughout the path, and a hyperadiabatic regime in which the error is given by a form similar to an asymptotic expansion in the inverse of the evolution time with the coefficients depending principally on the behavior at the endpoints. However, the error in this hyperadiabatic regime is neither given by a true asymptotic series nor solely dependent on the endpoints: the coefficients combine the contributions from both endpoints, with relative phase factors that depend on the average spectral gaps along the trajectory, multiplied by the evolution time. The central result of this paper is to identify a quantity, referred to as the typical error, which is obtained by appropriately averaging the error over evolution times that are small compared to the evolution time itself. This typical error is characterized by an asymptotic series and depends solely on the endpoints of the evolution, remaining independent of the details of the intermediate evolution.

\end{abstract}

\date{\today}
\maketitle


The adiabatic theorem~\cite{Born1928}, which emerged in the early days of quantum mechanics, has been influential in explaining various physical phenomena, such as the Born-Oppenheimer approximation~\cite{https://doi.org/10.1002/andp.19273892002} in molecular dynamics and topological phases~\cite{Vojta_2003, Hasan:2010xy, Sachdev:2011fcc}. 
Recently, the role of the adiabatic theorem has been emphasized in a direct application in quantum computing: namely, state preparation. For example, the adiabatic theorem underpins an approach to solving various optimization problems, where the solutions to the relevant issues are encoded in the ground state of the final Hamiltonian~\cite{APOLLONI1989233, farhi2000quantum, doi:10.1126/science.1057726, Childs:2001ge, Roland:2002huo, 959902, 1366223, doi:10.1137/060648829, RevModPhys.90.015002}.

In adiabatic quantum computation, the system Hamiltonian continuously, but slowly, evolves from an initial Hamiltonian $H_{i}$ (in which an eigenstate has been prepared at time $t_i$) to a final Hamiltonian $H_f$ (at time $t_f$). The success of this method relies on the fact that if the evolution of this Hamiltonian occurs infinitely slowly in time, i.e., $t_f-t_i\rightarrow\infty$, the system will remain in the corresponding eigenstate of the evolving Hamiltonian. 

The evolution is never infinitely slow in practice, and the finite rate of change of the Hamiltonian over time inevitably causes errors: there will always be transitions to other energy eigenstates of the instantaneous Hamiltonian. In the context of quantum computing, it is important to have an understanding of the errors induced by diabatic transitions. It turns out that the behavior of such errors is subtle. One approach, developed over the past two decades, is to find a rigorous upper bound for the timescale sufficient to ensure that the error is smaller than a fixed amount; such bounds depend on the minimum of the spectral gap and the norms of the derivatives of the Hamiltonian~\cite{10.1063/1.2798382, PhysRevA.80.012106, Cheung_2011, Mozgunov:2020cof, Burgarth2022oneboundtorulethem}. Unfortunately, this approach merely gives an upper bound rather than an estimate of the error. Moreover, there are deep reasons to believe that the upper bound is typically extremely loose and hence of limited utility.

An alternative approach is based on the so-called switching theorem, which gives a direct estimate for the error in the regime of asymptotically long times~\cite{GARRIDO1962553, 10.1063/1.1704127, FJSancho_1966, Nenciu1981, Avron1987, JP, Nenciu1991, Nenciu1993, HAGEDORN2002235,  10.1063/1.3236685, PhysRevA.82.052305, 10.1063/1.4748968, PhysRevLett.116.080503}.
The theorem describes the asymptotic behavior of the error accumulated as a function of evolution time, $T$, and the spectral gaps in the Hamiltonian. There is what might be termed a \textit{quasi}-asymptotic series for the error whose only dependence on the path between the initial and final endpoint are phase factors between a contribution from the initial and final points.

The switching theorem was discovered decades after the introduction of the adiabatic theorem~\cite{LENARD1959261} and has since been primarily studied within the field of mathematical physics.  
In the domain of its validity, the accuracy of the switching theorem improves as additional terms are included, up to a certain optimal number of terms, which generally depends on the value of $T$. However, if there are only a finite number of nonzero derivatives of the Hamiltonian at the endpoints, the series truncates at finite order. We will refer to the regime in $T$, where the asymptotic expansion is both effective and dominated by its leading term, as the hyperadiabatic regime. This regime lies within the adiabatic regime, where errors are small, and the adiabatic theorem approximately holds~\cite{doi:10.1143/JPSJ.5.435, messiah61,Tong:2005zz, PhysRevLett.104.120401}.



The evolution of a quantum state (using natural units with $\hbar \equiv 1$) is described by the time-dependent Schr\"odinger equation:
\begin{equation}
    i\frac{d}{dt} | \psi(t) \rangle = H(t) | \psi(t) \rangle. \label{Eq:SE}
\end{equation}
The adiabatic theorem implies that if $H(t)$ evolves infinitely slowly, then the $n$th eigenstate of the initial Hamiltonian evolves to the corresponding eigenstate of the final Hamiltonian. This paper focuses on the ground state, but, \textit{mutatis mutandis}, this can be done for any eigenstates. Since the system does not evolve infinitely slowly, diabatic transitions (errors) 
are unavoidable. Assuming $t_i$ and $t_f$ are the times at which the Hamiltonian corresponds to the initial and final Hamiltonians, the error accumulated during the evolution is given by
\begin{equation}
    \epsilon \equiv \left \lVert \left (1-|g_f\rangle \langle g_f |  \right) U(t_f, t_i)  |g_i \rangle \right \rVert .
    \label{Eq:errordef}
\end{equation}
In Eq.~(\ref{Eq:errordef}), $U(t_f, t_i)$ is the time evolution operator that evolves a quantum state at $t_i$ to another state at $t_f$ according to the Hamiltonian evolution for $H(t)$. $|g_i \rangle$, $|g_f \rangle$ are the ground states of the initial and final Hamiltonians, respectively.

In this work the time difference between the initial and final times is denoted by $T$, i.e.,  $T=t_{f}-t_{i}$. By an appropriate rescaling by $T$, we can re-express the  Schr\"odinger equation Eq.~(\ref{Eq:SE}) as follows:
\begin{equation}
    i\frac{d}{ds} | \psi(s) \rangle = T H(s) | \psi(s) \rangle .
\end{equation}
Here, $s \in [0,1]$. Note that $s$ is dimensionless. We will refer to $T$ as the timescale in this paper.

The switching theorem describes the propagation of errors that arises from the finiteness of the rate of Hamiltonian evolution in the regime of asymptotically long times. It says that the time evolution operator and the error follows a \textit{quasi}-asymptotic series\footnote{We call it a \textit{quasi}-asymptotic series since its coefficients depend on $T$ only through phase differences in contributions arising from the endpoints.} in $1/T$. More formally, the difference between the adiabatically evolved state and the desired ground state of the Hamiltonian can be expressed as the following \textit{quasi}-asymptotic series,
\begin{subequations}
\begin{align}
    U |g_i \rangle \langle g_i |U^\dagger - |g_f \rangle \langle g_f | &\, = \,\sum_{n=1}^{n_{\max}(T)}\frac{B_n}{T^n} + R, \label{Eq:seriesOp} 
\end{align}
where $U \equiv U(t_f, t_i)$, $B_n$ are operators that depend on $T$ only through phase factors. $R$, the remainder term, is an operator that scales with $T$ more slowly than any power of $T$. 
$n_{\max}(T)$ is the maximum number of terms for which this asymptotic series improves the accuracy of the description at $T$.
Equivalently, the error accumulated during the evolution has a similar \textit{quasi}-asymptotic series in $1/T$ as represented below;
\begin{align}
    \epsilon   &\, = \, \sum_{n=1}^{n_{\max}(T)} \frac{b_n}{T^n} + r,  \label{Eq:series}
\end{align}
\end{subequations}
$b_n$ and $r$ are numbers with properties analogous to those of $B_n$ and $R$. The operators $B_n$ and the coefficients $b_n$ depend principally on the properties of the endpoints of the trajectory; the only effect of the trajectory between the endpoints is to introduce phase factors in differences between the endpoint contributions.
Moreover, the endpoint contributions for $b_n$ depend only on the matrix elements of the $n$th derivative of $H$ with respect to the scaled time.

This implies that if only a finite number of derivatives are nonzero, then there will only be a finite number of nonzero coefficients and the series will truncate. For asymptotically large values of $T$, the switching theorem implies that up to oscillations due to the phase factors, the errors behave as $(1/T)^{k}$ where $k$ is the smallest integer for which $b_{k}$ is non-zero. 

The $b_n$ coefficients can be calculated perturbatively~\cite{PhysRevA.73.042104, PhysRevA.78.052508, PASSOS2020168172, Cohen:2024nbk}. In typical cases where the first derivative of the Hamiltonian with time at either the initial or the final point is non-zero, the \textit{quasi}-asymptotic series is dominated by the $1/T$ term for large evolution times. The coefficient $b_1$ is given by
\begin{widetext}
\begin{equation}
\begin{aligned}
    b_1 & = \sqrt{  \sum_{j \neq g} \left| {\rm e}^{i w_{j, g} T} \frac{ \langle j(1) | H'(1) | g(1) \rangle}{\Delta_{j, g}^2(1)} - \frac{ \langle j(0) | H'(0) | g(0) \rangle}{\Delta_{j, g}^2(0)} \right|^2 },
\end{aligned} \label{Eq:b1}
\end{equation}
\end{widetext}
where $\Delta_{j, g}(t) = E_j(t) - E_g(t)$ and $w_{j, g} = \int_0^1 ds \, \Delta_{j, g}(s)$. Thus $w_{j, g}$ is the average spectral gap between the ground state and the $j$th excited state, calculated as an average over the trajectory. The sum over $j$ is over all excited states of the Hamiltonian. The expression for $b_1$ indicates that if $H'(s)$ is zero at the endpoints, $b_1$ vanishes, causing the error to scale as $1/T^2$ in the asymptotic limit (using Eq.~(\ref{Eq:series})). More generally, when $H^{(m)}(0)=H^{(m)}(1)=0$ for all $m \le n$ and $H^{(n)} \ne 0$ at least one endpoint (where $H^{(n)}$ is the $n$th derivative with respect to the scaled time), the coefficient $b_n$ is given by
\begin{widetext}
\begin{equation}
\begin{aligned}
    b_n & = \sqrt{  \sum_{j \neq g} \left| {\rm e}^{i w_{j, g} T} \frac{ \langle j(1) | H^{(n)}(1) | g(1) \rangle}{\Delta_{j, g}^{n+1}(1)}- \frac{ \langle j(0) | H^{(n)}(0) | g(0) \rangle}{\Delta_{j, g}^{n+1}(0)} \right|^2} \; ;
\end{aligned} \label{Eq:bn}
\end{equation}
\end{widetext}
in these circumstances, the error qualitatively scales asymptotically as $b_{n}/T^{n}$ in the hyperadiabatic regime.


However, note that the coefficient $b_n$ depends on $T$ through the phase factors ${\rm e}^{i w_{j, g} T}$. Thus, the scaling of the error with respect to $T$ is rather complicated, despite the apparent simple qualitative scaling as $1/T^n$. Moreover, the phase factors mean that the error does not depend solely on the endpoints but also on the behavior of the system along the path.

\begin{figure}[t]
    \includegraphics[width=0.49\textwidth]{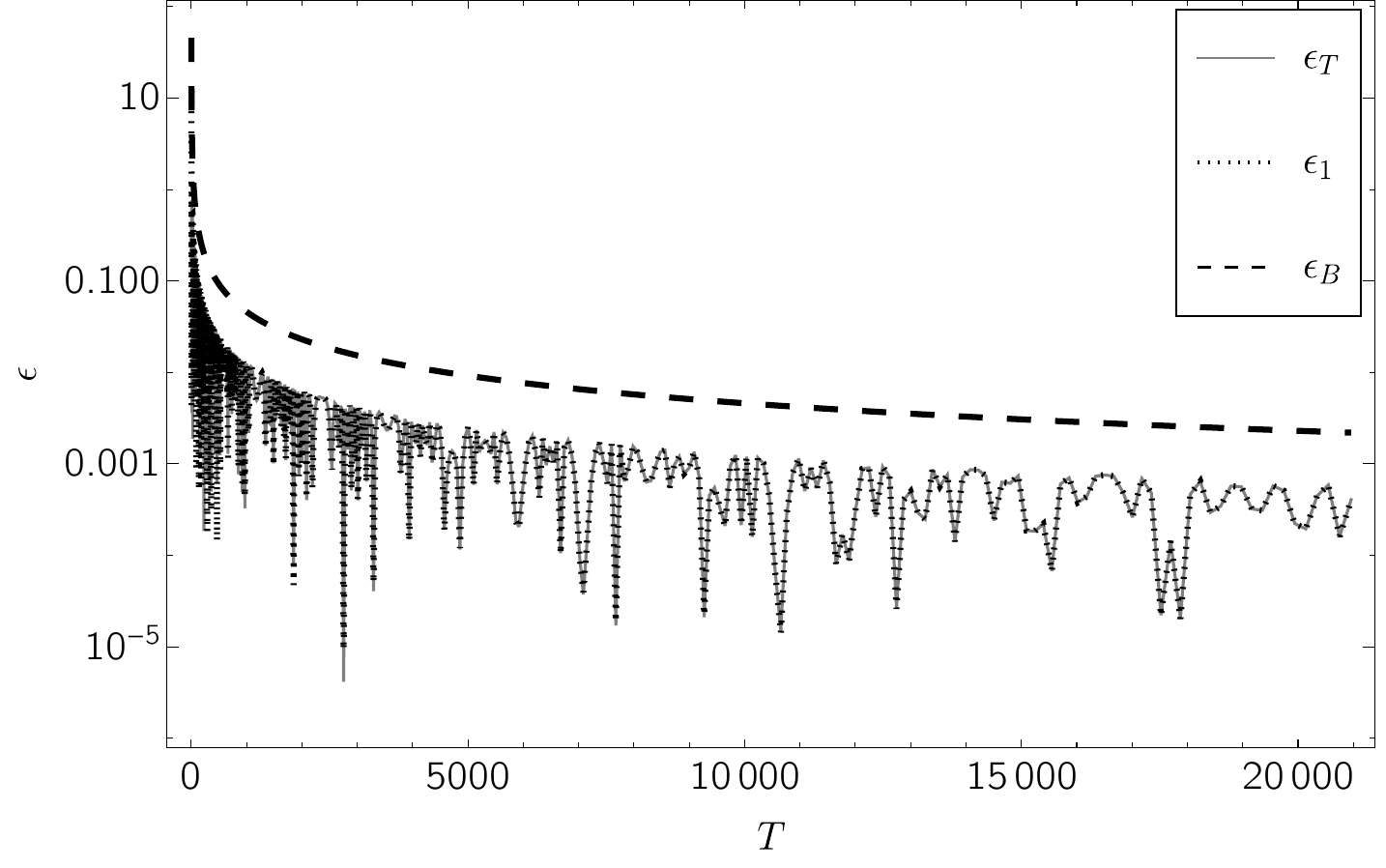}
    \caption{
    Comparison of the true error ($\epsilon_T$, solid), switching error ($\epsilon_1$, dotted), and the upper bound of the error using the norms of the Hamiltonian's derivatives and the minimal spectral gap ($\epsilon_B$, dashed) from~\cite{10.1063/1.2798382}, for the Hamiltonian in Eq.~(\ref{Eq:H0}).
    }
    \label{fig1}
\centering 
\end{figure}

All the challenges associated with estimating errors using the upper bound, which is based on the norms of the Hamiltonian's derivatives with the minimal spectral gap, along with the switching theorem, as discussed above, are illustrated in Fig.~\ref{fig1}. It compares different estimates of errors with the true error, $\epsilon_T$, as defined in Eq.~(\ref{Eq:errordef}), by directly solving the Schr\"odinger equation. The Hamiltonian is given as
\begin{equation}
H(s) = 
\begin{bmatrix}
 s(1-s) & 0.2  \\
 0.2 & -s(1-s)
\end{bmatrix},  \label{Eq:H0}
\end{equation}
where $s\in[0,1]$. Note that $H'(s)$ is nonzero at the endpoints and therefore the asymptotic switching error $\epsilon_1$ is defined as $b_1/T$ where $b_1$ is in Eq.~(\ref{Eq:b1}). $\epsilon_B$ is defined from Ref.~\cite{10.1063/1.2798382}, which found the upper bound of errors using the minimum spectral gap and the norms of the derivatives of the Hamiltonian. Fig.~\ref{fig1} demonstrates that the upper bound of errors highly overestimates the true error. While the switching error $\epsilon_1$ follows the true error very precisely at large timescales, both exhibit significant fluctuations, which is inherent to the adiabatic theorem. Consequently, it becomes challenging to predict errors in realistic situations, where the exact determination of the timescale $T$ is rarely feasible.

This paper explores how these issues can be avoided with a slight change of focus on a quantity that differs from the asymptotic expression for the error itself. The goal is to identify a quantity associated with the typical value of the error that has the properties of exhibiting a simple power-law scaling with $T$ in the hyperadiabatic regime, and whose value depends solely on the endpoints. This can be achieved by considering the typical value of the error for timescales in the vicinity of $T$ by averaging the error over a range of $T$ values. The range should be much larger than the inverse of $w_{1, g}$ (the average spectral gap between the ground state and the lowest excited state) but much smaller than $T$.

Formally, one can define the ``typical error'' for a (long) timescale $T$, $\bar\epsilon(T)$, as 
\begin{equation}
\bar\epsilon(T) \equiv  \frac{1}{2 \sqrt{T \tau_0}} \int_{T-\sqrt{T \tau_0}}^{T+\sqrt{T \tau_0}} d T' \,\epsilon(T')    \label{Eq:barerror}
\end{equation}
for some (arbitrary) positive choice of $\tau_0$. As $T$ becomes large, every choice of $\tau_0$ yields an average over a range of time scales that is large compared to the scale of fluctuations in $\epsilon$ and small compared to $T$ itself, producing a result independent of $\tau_0$.
In cases where the lowest non-vanishing derivative of the Hamiltonian at the endpoints is $n$, Eq.~(\ref{Eq:bn}) implies that in the hyperadiabatic regime, $\bar\epsilon(T)$ is given by
\begin{widetext}
\begin{equation}
    \bar\epsilon (T) = \frac{\bar b_n}{T^n} \; \; {\rm with} \; \;
    \bar{b}_n  =  \sqrt{ \sum_{j \neq g}   \left| \frac{ \langle j(0) | H^{(n)}(0) | g(0) \rangle}{\Delta_{j, g}^{n+1}(0)} \right|^2 + \sum_{j \neq g}\left| \frac{ \langle j(1) | H^{(n)}(1) | g(1) \rangle}{\Delta_{j, g}^{n+1}(1)} \right|^2  } 
    \equiv \sqrt{(\bar{b}_n^{0})^2 + (\bar{b}_n^{1})^2}  . \label{Eq:barbn}
\end{equation} 
\end{widetext}

There are several virtues to using the typical error, $\bar{\epsilon}(T)$, rather than $\epsilon (T)$ itself. Two obvious motivations for the introduction of $\bar{\epsilon}(T)$ are: first, the fact that asymptotically $\bar{\epsilon}(T)$ has a simple power-law scaling in $T$, and second, that its value depends solely on the behavior at the endpoints, making the asymptotic behavior completely independent of the details of how the path is traversed. Indeed, the mean square of $\epsilon$, $\bar{\epsilon^2}(T)$, is given by $\left((\bar{b}_n^{0})^2 + (\bar{b}_n^{1})^2\right)/T^{2n}$, which is the sum of two contributions---one coming entirely from the initial point and the other from the final point.

However, this quantity has several additional virtues. In practice, one may not be able to control the traversal with arbitrary accuracy---and even if one could, one may well not have sufficient knowledge of the intermediate path to determine the phase factors. In such cases, the typical error, $\bar \epsilon (T)$, provides something akin to the expected value of the error. Moreover, in certain cases, $\bar{\epsilon}(T)$ can be used to estimate the error directly, as in these cases, $\epsilon(T)$ is very close to $\bar{\epsilon}(T)$ for almost all asymptotic $T$. This occurs when the significant contributions to $b_n$ in Eq.~(\ref{Eq:bn}) are spread over a large number of excited states so that the phase factors in the various terms essentially cancel. Furthermore, the typical error can be used to establish an upper bound of the error in the hyperadiabtic regime. From the structures of Eq.~(\ref{Eq:bn}) and Eq.~(\ref{Eq:barbn}) it is apparent that in the hyperadiabatic regime
\begin{equation}
    \epsilon(T) \le   \sqrt{2} \bar\epsilon(T) . \label{Eq:ub}
\end{equation}
The ability to find an upper bound of the error is potentially of value in the context of quantum computing. That said, the upper bound on the error in  Eq.~(\ref{Eq:ub}) is not a true upper bound for any $T$; it only serves as an upper bound if $T$ is large enough for the system to be in the hyperadiabatic regime, where the error is dominated by the leading term in the asymptotic series\footnote{Before entering the hyperadiabatic regime, the upper bound of the error may depend on the path through a measure known as the path length, denoted as $L \equiv \int_0^1 ds \, \left\lVert \frac{d}{ds}|g(s) \rangle - |g(s)\rangle \langle g(s)| \frac{d}{ds} | g(s) \rangle \right\rVert$~\cite{10.5555/2011804.2011811, boixo2010fast, PhysRevA.89.012314, Cohen:2023dll, Cohen:2024nbk, Cohen:2024kdi}.}.

\begin{figure}[t]
    \includegraphics[width=0.49\textwidth]{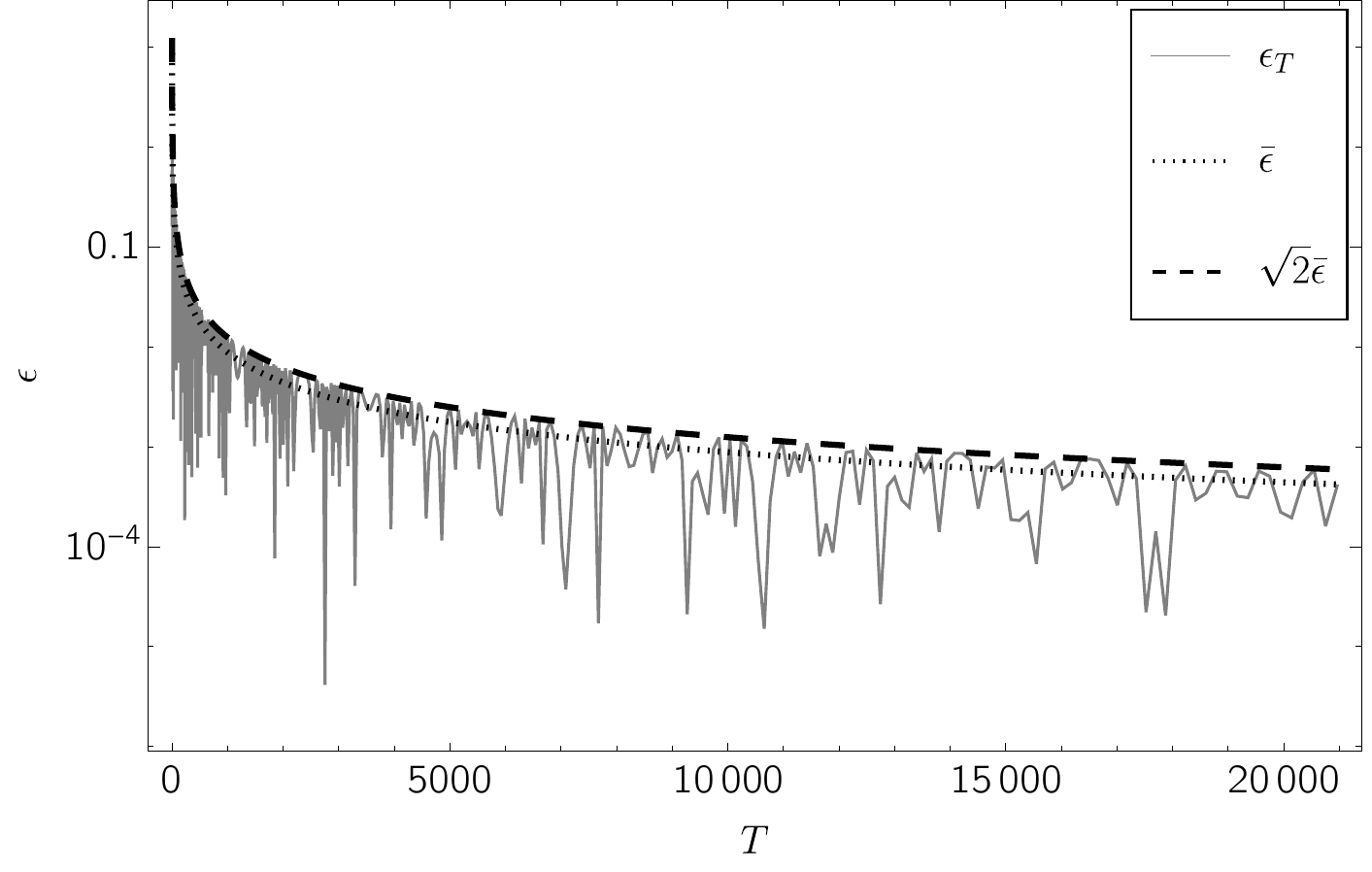}
    \caption{
    Comparison of the true error ($\epsilon_T$, solid), typical error ($\bar{\epsilon}$, dotted), and the upper bound of the error in the hyperadiabatic regime ($\sqrt{2}\bar{\epsilon}$, dashed), for the Hamiltonian in Eq.~(\ref{Eq:H0}).
    }
    \label{fig2}
\centering 
\end{figure}

Fig.~\ref{fig2} compares the true error, $\epsilon_T$, with the typical error, $\bar{\epsilon}$, with its bound, $\sqrt{2}\bar{\epsilon}$. It shows that the typical error avoids the fluctuations seen in the actual error.  It also illustrates that in the hyperadiabatic regime, the upper bound from the typical error matches the upper bound in the fluctations of the actually error extremely well.

In summary, this paper demonstrated that when the system is in the hyperadiabatic regime, a natural quantity to consider is the typical error defined in Eq.~(\ref{Eq:barbn}) which has a well-defined power-law scaling in $1/T$ and is insensitive to the details of the traversal of the path except at its endpoints. A key issue in the context of adiabatic quantum state preparation is the extent to which it is necessary to be in the hyperadiabatic regime in order to get usefully small errors. This issue will be addressed in a forthcoming paper.

\begin{acknowledgments}

This work was supported in part by the U.S. Department of Energy, Office of Nuclear Physics under Award Number(s) DE-SC0021143, and DE-FG02-93ER40762.

\end{acknowledgments}

\bibliography{refs.bib}

\begin{widetext}

\appendix
\section{A loose derivation of the typical error for the switching error formula}

While the typical error is formally defined in Eq.~(\ref{Eq:barerror}), one can expect its form by considering the square of Eq.~(\ref{Eq:bn}):
\begin{equation}
\begin{aligned}
    b_n^2  =  \sum_{j \neq g} & \Bigg(  \left| \frac{ \langle j(1) | H^{(n)}(1) | g(1) \rangle}{\Delta_{j, g}^{n+1}(1)} \right|^2 + \left |\frac{ \langle j(0) | H^{(n)}(0) | g(0) \rangle}{\Delta_{j, g}^{n+1}(0)} \right|^2 \\
    & + {\rm e}^{i w_{j, g} T}  \frac{ \langle j(1) | H^{(n)}(1) | g(1) \rangle}{\Delta_{j, g}^{n+1}(1)} \frac{ \langle g(0) | H^{(n)}(0) | j(0) \rangle}{\Delta_{j, g}^{n+1}(0)} 
    + {\rm e}^{-i w_{j, g} T}  \frac{ \langle g(1) | H^{(n)}(1) | j(1) \rangle}{\Delta_{j, g}^{n+1}(1)} \frac{ \langle j(0) | H^{(n)}(0) | g(0) \rangle}{\Delta_{j, g}^{n+1}(0)} 
    \Bigg) \, .
\end{aligned}
\end{equation}
On average, the last two terms oscillate rapidly and vanish, leaving the typical error to follow Eq.~(\ref{Eq:barbn}). There can be alternative definitions of the typical error, such as using a root-mean-square with timescale averaging. However, these alternatives will lead to the same formula as Eq.~(\ref{Eq:barerror}) when the timescale-dependent terms are averaged out.

\end{widetext}

\end{document}